\documentclass[runningheads,a4paper]{llncs}

\usepackage{amssymb}
\usepackage{graphicx}
\usepackage{tcolorbox}
\usepackage{multirow}
\usepackage{subcaption}
\usepackage{url}
\begin{document}

\mainmatter  

\title{Are Non-Experts Able to Comprehend \\ Business Process Models - Study Insights Involving Novices and Experts}

\titlerunning{Are Non-Experts Able to Comprehend Business Process Models?}
\author{Michael Winter\inst{1}, R{\"u}diger Pryss\inst{2}, Thomas Probst\inst{3}, Winfried Schlee\inst{4}, Miles Tallon\inst{5}, Ulrich Frick\inst{5}, \and Manfred Reichert\inst{1}}

\institute{Inst of Databases and Information Systems, Ulm University, Germany
	\and
	Inst of Clinical Epidemiology and Biometry, University of Würzburg, Germany
	\and
	Dep for Psychotherapy and Biopsycho Health, Danube University Krems, Austria
	\and
	Dep of Psychiatry and Psychotherapy, Regensburg University, Germany
	\and
	Hochschule D{\"o}pfer, University of Applied Sciences, Germany
	\\
	\{michael.winter, manfred.reichert\}@uni-ulm.de, ruediger.pryss@uni-wuerzburg.de, thomas.probst@donau-uni.ac.at, winfried.schlee@googlemail.com, \\ \{m.tallon, u.frick\}@hs-doepfer.de}

\authorrunning{Zimoch et al.}
%
%

\maketitle

\begin{abstract}

The comprehension of business process models is crucial for enterprises. Prior
research has shown that children as well as adolescents perceive and interpret
graphical representations in a different manner compared to grown-ups. To
evaluate this, observations in the context of business process models are
presented in this paper obtained from a study on \textit{visual literacy in
cultural education}. We demonstrate that adolescents without expertise in
process model comprehension are able to correctly interpret business process
models expressed in terms of BPMN 2.0. In a comprehensive study, $n=205$
learners (i.e.,  pupils at the age of 15) needed to answer questions related to
process models they were confronted with, reflecting different \textit{levels of
complexity}. In addition, process models were created with varying styles of
element labels. Study results indicate that an \textit{abstract} description
(i.e., using only alphabetic letters) of process models is understood more
easily compared to \textit{concrete} or \textit{pseudo} descriptions. As
benchmark, results are compared with the ones of modeling experts ($n=40$).
Amongst others, study findings suggest using \textit{abstract} descriptions in
order to introduce novices to process modeling notations. With the obtained
insights, we highlight that process models can be properly comprehended by
novices.

\end{abstract}

\section{Introduction}

\textit{Business process models} constitute crucial artifacts for optimizing the
operational efficiency of enterprises. The demand for process models of high
quality, which properly capture the business processes of an enterprise, has
increased during the last years \cite{van2016business}. To assist enterprises in
meeting this demand, considerable research has been spent on better
understanding these factors that characterize process models of high quality
\cite{becker2000guidelines,DBLP:books/sp/Krogstie16}. However, as a prerequisite
for an effective use of process models, the latter must be properly understood
by individuals \cite{figl2017comprehension}.   \\
Concerning \textit{process model comprehension}, an individual must parse
information related to the \textit{syntactics}, \textit{semantics}, and
\textit{pragmatics} of a process modeling notation
\cite{lindland1994understanding}. Prior research has shown that individuals
apply different strategies for interpreting and comprehending process models
\cite{zimoch2}. In this context, it is known that children, adolescents,
and grown-ups perceive their surroundings differently. As a consequence, they
use varying strategies for interpreting and learning artifacts
\cite{thomas2004evidence,brehmer2008comparing}. In certain cases, children and
adolescents show an equivalent or even better performance in accomplishing
cognitive tasks compared to grown-ups
\cite{opfer2008representational,lucas2014children}. This raises the issue
whether children and adolescents are also able to comprehend business process
models, even though they have no previous knowledge in any process modeling
notation. Following this, first, we believe that from corresponding insights we
can draw conclusions fostering the comprehension of process models. Second, modeling
guidelines (e.g., 7PMG \cite{Mendling2010}) towards creating better
comprehensible process models might be derived. Third, BPM modeling tools can be
augmented with features to foster the learning of process modeling notations and,
thus, the comprehensibility of business process models. \\
This paper presents the results of a comprehensive study on \textit{visual
literacy in cultural education}. More specific, a representative sample ($n =
205$) of \textit{pupils at the age of 15} from different kinds of German schools
are confronted with a visual task related to process model comprehension.
Particularly, the pupils have no previous knowledge on any process modeling
notation. The objective of the study is to evaluate whether pupils (i.e.,
novices) comprehend process models correctly. Study results imply that pupils
are able to comprehend process models. Further, as a benchmark, a similar study
is conducted with participants having expertise in the domain of process
modeling. Based on the study results, we give recommendations on how to foster
process model comprehension.   \\
The remainder of the paper is organized as follows: Section 2 explains the study
context and Section 3 the study settings. Study results, in turn, are presented,
analyzed, and discussed in Section 4. Section 5 addresses related work. Finally,
Section 6 summarizes the paper and gives an outlook.

\section{Study Context}
The presented results are obtained from a large-scale study in a project
focusing on the \textit{visual literacy in cultural
education}.\footnote[1]{\scriptsize{}\url{https://www.dipf.de/en/research/current-projects/bkkb-visual-literacy-in-cultural-education?set_language=en}}
The purpose of this project is to gain insights into how \textit{visual
literacy} can be fostered and empirically measured \cite{tallon}. Thereby,
visual literacy denotes a concept defining the capability to interpret,
understand, and extract information presented in images \cite{stokes2002visual}.
Furthermore, studies have shown that the use of appropriate images foster
learning processes as well as the development of learning strategies
\cite{tallon2}. The project analyzes the variation between motivation and
perception of learners (e.g., pupils, students) about instructional quality with
regard to visual literacy in cultural education. A particular emphasis is put on
the effects of \textit{cultural capital} (i.e., social assets like education)
and \textit{cultural-aesthetic practices} (e.g., religious beliefs) of learners.
Cultural education will be assessed in the \textit{European Framework of Visual
Literacy (ENViL)} in a series of nationwide studies across various kinds of
schools (e.g., secondary school, university) in Germany \cite{envil}. For this
purpose, a mobile application was developed to assess visual literacy. In more
detail, learners \raisebox{.5pt}{\textcircled{\raisebox{-.9pt} {1}}} are asked
to run through various \textit{visual tasks} (e.g., Dalli-Click, Mental
Rotation) and a set of questions \cite{greenfield2009technology}. Additional
data are collected through eye tracking to reveal insights into learning
processes and strategies (cf. Fig. 1). \\
To the aforementioned study, we contribute a visual task that deals with the
comprehension of process models. In particular, the focus lies primarily on
\textit{semantics} and \textit{syntactics} of process models expressed in
\textit{BPMN 2.0} \cite{aagesen2010analysis,OMG2010}. With this task, we want to
investigate the following objectives with respect to process model
comprehension:

\begin{tcolorbox}[title=Objectives for Process Model Comprehension]
\small
\begin{itemize}
	\item Can process model semantics be interpreted by learners? 
	\item Can process model syntactics be comprehended by learners?
	\item Does the interpretation and comprehension of process models change when taking cultural education into account?  
\end{itemize}
\end{tcolorbox}

The study results, in turn, can be used to discover directives on how to foster
process model comprehension. Moreover, the introduction of modeling notations to
novices might be improved and BPM tools could be enhanced with features
fostering process model comprehension. As a benchmark, we conduct a similar
study with participants experienced with business process modeling 
\raisebox{.5pt}{\textcircled{\raisebox{-.9pt} {2}}}.
In detail, instead of carrying out the entire visual literacy study as for
pupils, including all visual tasks and questions, in the
benchmark study, participants only need to answer questions on their process
modeling experience. Furthermore, they only solve the process model
comprehension task. Fig. 1 summarizes the described study context of visual
literacy.

\begin{figure}
	\centering
	\includegraphics[width=.7\linewidth]{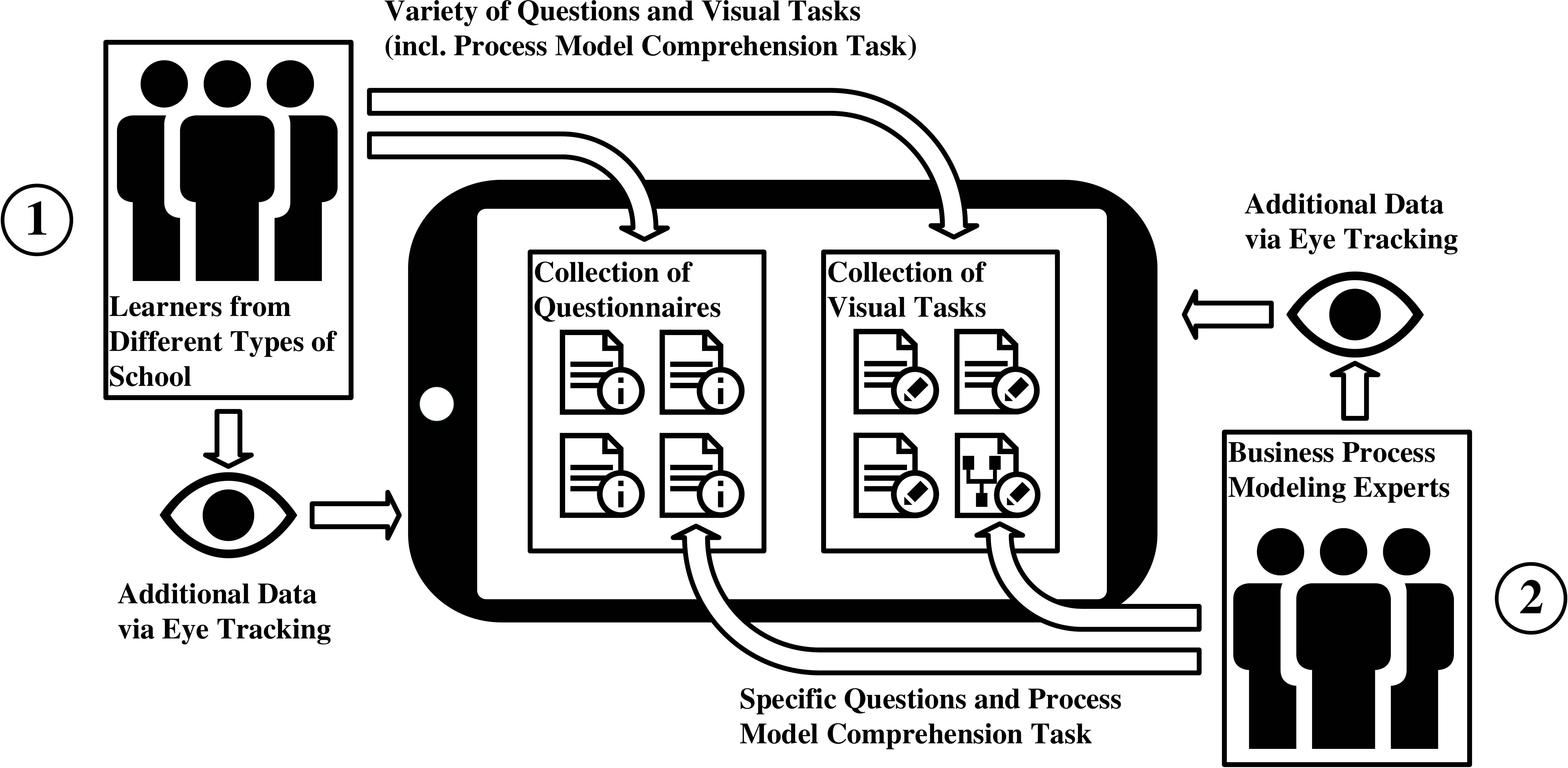}
	\caption{Visual Literacy in Cultural Education}
	\label{fig:vis}
\end{figure}

\section{Study Setting}
In general, there are many factors that have effects on the comprehension of
process models \cite{Mendling2012}. Amongst others, the size of a process model
(i.e., \textit{\textit{level of complexity}}) and the quality of element labels
are such factors \cite{leopold2012refactoring,latva2001finding}. The latter is
relevant for the semantic description of process models, which comprises the
textual as well as informational content. In practice, the use of
\textit{concrete labels} (e.g., verb-object style) is common. However,
\textit{abstract labels} (i.e., alphabetic letters) may be used, if the semantic
description of a process model is not relevant, e.g., when getting into touch
with process modeling notations for the first time. In the study, we introduce an
additional labeling style for elements using \textit{pseudo labels}, i.e., we
generate appropriate pseudowords for all nouns, derived from the concrete
labeling style. The different labeling styles (i.e., \textit{concrete},
\textit{abstract}, and \textit{pseudo}) are denoted as \textit{levels of
semantics} in the following. Based on these two factors (i.e., \textit{level of
complexity} and \textit{level of semantics}), the following three research
questions (RQ 1 - RQ 3) are addressed with learners pupils in the study:

\begin{center}
	\small
	\begin{tabular}{p{1.2cm}p{10.8cm}}
		\hline
		\textbf{RQ 1:} & \textit{How do pupils (i.e., novices) perform when reading and comprehending BPMN 2.0 process models of different levels of complexity?}  \\ \hline
		\textbf{RQ 2}: & \textit{How do pupils (i.e., novices) perform when reading and comprehending BPMN 2.0 process models of different levels of semantics?} \\ \hline
		\textbf{RQ 3:} & \textit{How do pupils (i.e., novices) perform with respect to the reading and comprehension of BPMN 2.0 process models compared to process modeling experts?} \\ \hline
	\end{tabular}
\end{center}

%
%

\subsection{Study Planning}
\label{setup}


\textbf{Participants.} Study participants form two samples. The first sample
comprises \textit{pupils at the age of 15} from different kinds of German
schools. The second sample, in turn, either refers to \textit{students} or
\textit{research associates} (i.e., process modeling experts) at Ulm University.
The latter are invited separately for participating in the benchmark study. As a
prerequisite for participating in the study, benchmark participants need to have
a sufficient expertise level in process modeling. \\
\textbf{Objects.} The objects include three process models expressed in terms of
\textit{BPMN 2.0}. As being frequently used in practice, BPMN
2.0 suits to the context of this research. The process models are divided
into three \textit{levels of complexity} (i.e., \textit{easy}, \textit{medium},
and \textit{hard}). The \textit{easy} process model only comprises a sequence
of basic elements (i.e., activities, events). With rising \textit{level of
complexity}, new elements, previously not contained in the process model, as
well as specific modeling constructs (e.g., loop) are added and the total number
of elements is increased. \\
Moreover, the process models represent different scenarios the participants have
experienced repeatedly in their daily lives, i.e., taking the bus home, browsing
Facebook while listening to music, and writing an exam. \\
For each \textit{level of complexity}, the corresponding process models reflect
three \textit{levels of semantics} (i.e., \textit{concrete}, \textit{abstract},
and \textit{pseudo}). For creating the latter, we use the multilingual
pseudoword generator \textit{Wuggy} \cite{keuleers2010wuggy}.  Figs. 2 a - c
illustrate the labeling styles corresponding to the three \textit{levels of
semantics}.  \\
For each process model, four statements on its semantics are presented to the
participants, who then need to answer which of the four statements are correct.
More precisely, two of the four statements are true, whereas the two others are
false (i.e., \textit{two-out-of-four} combination). Thereby, no information is
given to the participants about the two-out-of-four combination. The statements
are used to evaluate whether or not the participants interpret the process
models correctly. For collecting answers, check boxes are placed beneath each
statement, i.e., participants can easily select or deselect statements (cf.
Sect. 3.3). Initially, all statements are deselected. \\
To enable a comparability of the different process models, process modeling
experts as well as novices, who do not participate in the study, are asked to
rank and categorize the process models with respect to their \textit{level of
complexity} and \textit{level of semantics}. Finally, a steady increase in the
\textit{level of complexity} is ensured by applying quality metrics for process
models\footnote[2]{\scriptsize{}Material download at:
\url{https://drive.google.com/open?id=1ODBzR1eS0Tv5hhNyf6A1ai-R39HVH6Ks}}
\cite{vanderfeesten2007quality}. \\
\textbf{Independent variables.} The study comprises three independent variables,
i.e., for each process model, the \textit{level of complexity} (i.e.,
\textit{easy}, \textit{medium}, and \textit{hard}), the \textit{level of
semantics} (i.e., \textit{concrete}, \textit{abstract}, and \textit{pseudo}),
and, only for the benchmark study, the \textit{level of expertise} on process
modeling of experts.  \\
\textbf{Dependent variables.} Dependent variables are the \textit{achieved
score} regarding the statements and the \textit{duration needed} for
comprehending a process model.\\

\begin{figure}[]
	\centering
	\begin{subfigure}{0.20\textwidth}
		\includegraphics[width=\linewidth]{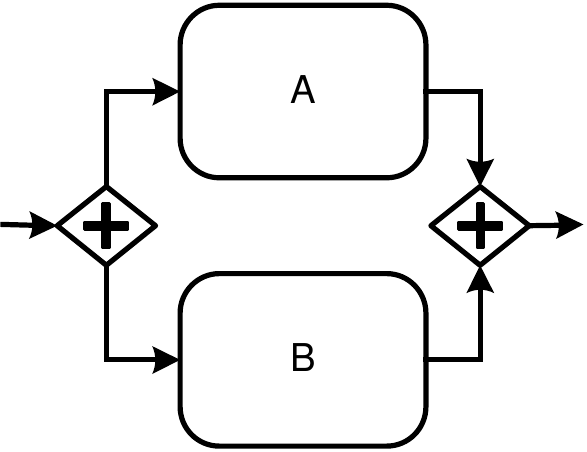}
		\caption{Abstract}
	\end{subfigure}
	\begin{subfigure}{0.20\textwidth}
		\includegraphics[width=\linewidth]{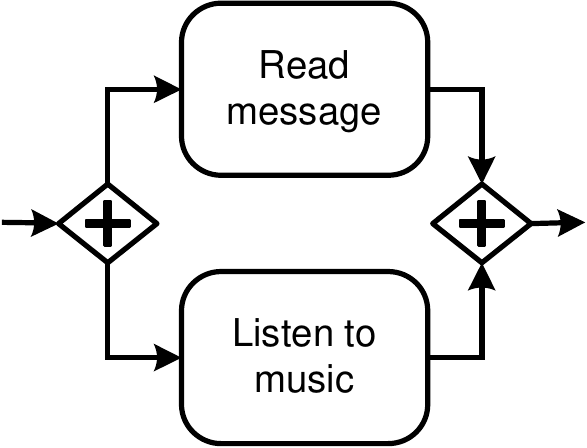}
		\caption{Concrete}
	\end{subfigure}
	\begin{subfigure}{0.20\textwidth}
	\includegraphics[width=\linewidth]{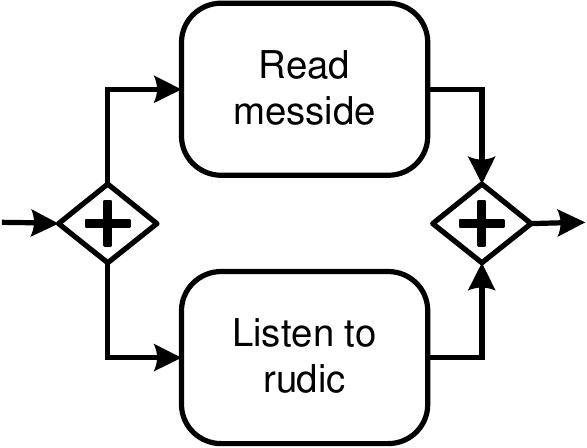}
	\caption{Pseudo}
\end{subfigure}
\caption{Levels of Semantics}
\end{figure}

\subsection{Study Design and Procedure}
Prior to the primary study, a pilot study with 17 students was performed to
evaluate the used process models and statements as well as to eliminate
misunderstandings and ambiguities. The pupils are told that the
study contains process model comprehension tasks. In turn, the experts are
informed that the study deals with process model comprehension. Both samples are
asked to perform the task of the study as quickly as possible and as careful as
possible. \\
As described in Sect. 2, there are two different procedures for pupils and
experts during the execution of the study (cf. Fig. 3). Pupils
\raisebox{.5pt}{\textcircled{\raisebox{-.9pt} {1}}} need to work on a predefined
sequence of visual tasks related to visual literacy and answer a number of
questions providing personal information (e.g., age, gender) and cultural
capital (cf. Sect. 2). Regarding the process model comprehension task, a
corresponding description is displayed on the tablets that explains what needs
to be done. In particular, the pupils are asked to read and comprehend the
depicted process models, starting with the process model reflecting an
\textit{easy} \textit{level of complexity}, followed by the process model with
\textit{medium level of complexity}, and the one with \textit{hard level of
complexity}. Thereby, each pupil is randomly assigned to a specific
\textit{level of semantics} (i.e., \textit{concrete}, \textit{abstract}, or
\textit{pseudo}) such that all elements in the three process models use the same
element labeling style. \\
Regarding the experts  \raisebox{.5pt}{\textcircled{\raisebox{-.9pt} {2}}},
general study information is presented, followed by a demographic questionnaire
(e.g., age, gender). A particular focus is put on questions related to the
expert's present knowledge on process modeling and the number of process models
he or she has analyzed and created during the last 12 months. After completing
this mandatory step, experts are confronted with the same task on process model
comprehension as the pupils. While the pupils have to solve and answer
additional visual tasks and questions, for the experts the study ends after
completing the respective comprehension task. Fig. 3 illustrates the two study
designs for pupils and experts.

\begin{figure}
	\centering
	\includegraphics[width=.7\linewidth]{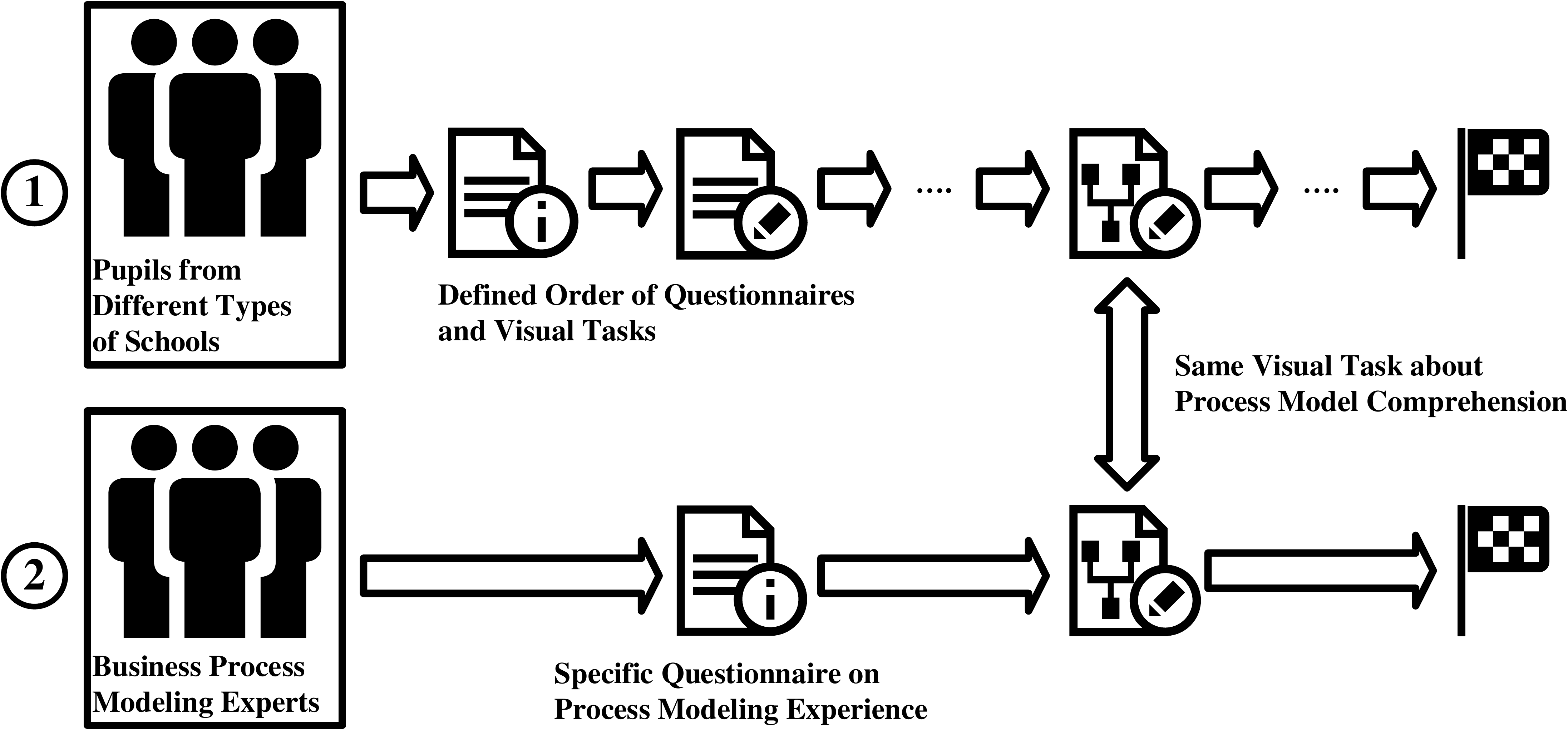}
	\caption{Study Design for Pupils and Experts}
	\label{fig:app}
\end{figure}

\subsection{Instrumentation}
The study is performed using a Samsung Galaxy Tab A6. Therefore, we developed a
mobile application, which serves as an instrument allowing for the planning and
execution of studies in the context of visual literacy. It comprises a variety
of customizable questions as well as visual tasks (e.g., Dalli-Click, Mental
Rotation). Emerging study data is collected with the mobile application. Fig. 4 
illustrates the user interface of the mobile application, showing the screen
when performing the visual task related to process model comprehension. Finally,
IBM SPSS Statistics 23 is used for all statistical analyses.


\begin{figure}[]
		\centering
		\includegraphics[width=0.65\linewidth]{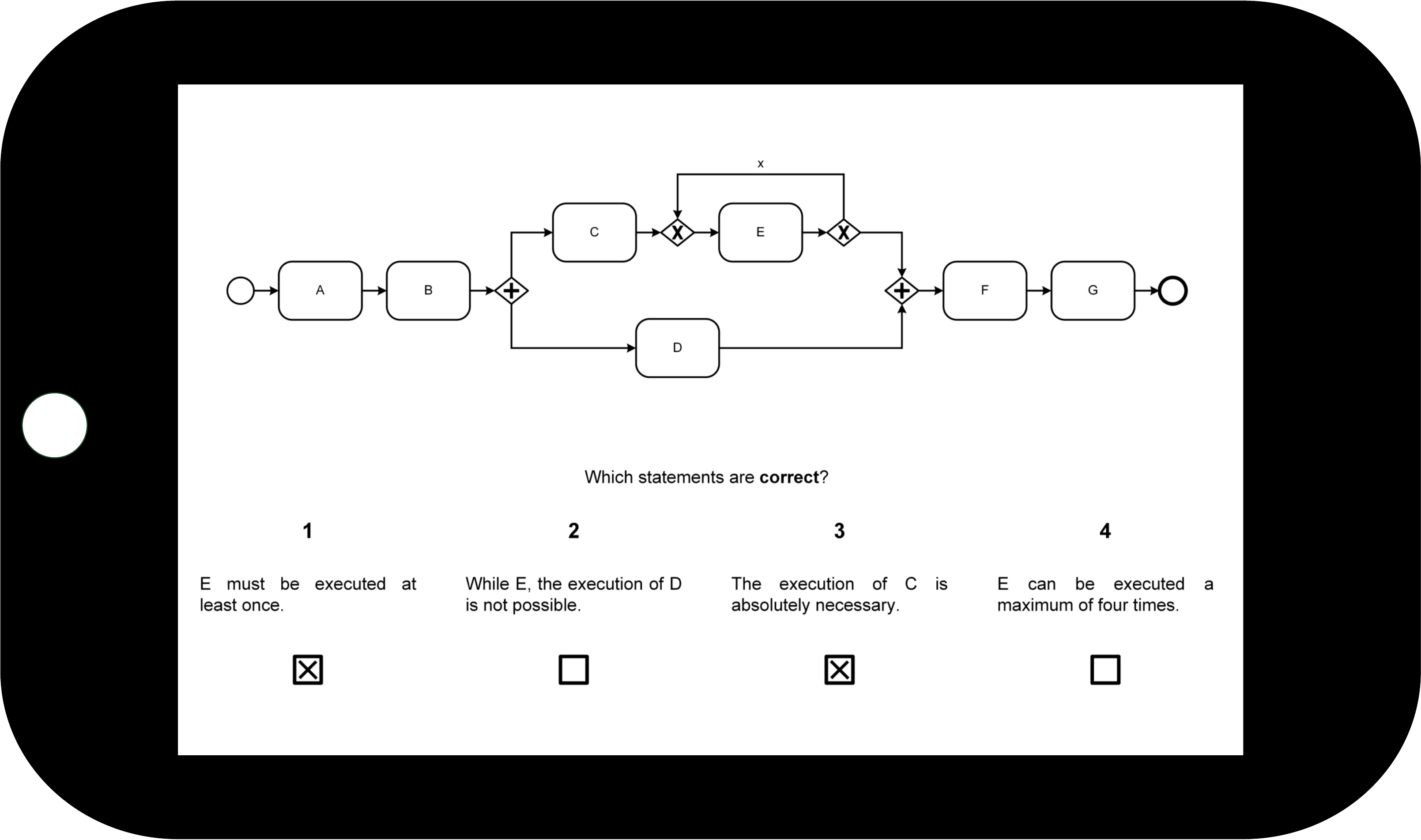}
		\caption{User Interface of the Mobile Application}
\end{figure}

\section{Data Analysis and Interpretation}
A total of 268 participants are recruited for both studies, forming two samples.
The first one consists of 228 learners (i.e., pupils, students) from different
types of German schools. As solely pupils are considered, data sets produced by
other learners are removed (i.e., 23 in total), leaving $n=205$ data sets left
for statistical analyses. The second sample, in turn, consists of students and
research associates from Ulm University, stemming from various departments
(e.g., Computer Science, Economics). As a prerequisite for their selection, they
must have a sufficient expertise level in business process modeling.
Accordingly, this sample is classified as \textit{process modeling experts} and
used as a benchmark for evaluating the results obtained from the study with
pupils. Finally, the stated research questions (cf. Sect. 3.1) are investigated
for \textit{$n=205$ pupils (i.e., novices) around the age of 15} and
\textit{$n=40$ modeling experts} respectively. Table 1 summarizes the sample
descriptions, displaying general information (with \textit{standard deviation
(SD)}) on the total \textit{number of participants}, their \textit{average age},
and \textit{gender balance}. Concerning process modeling experience, the average
number of process models \textit{analyzed} and \textit{created} during the last
12 months is shown as well as the average \textit{number of activities} in these
process models. Regarding the novices, no such data is available (NA) as they
have no experience in process modeling and process model comprehension
respectively. Finally, for each \textit{level of semantics} (i.e.,
\textit{concrete}, \textit{abstract}, and \textit{pseudo}), its distribution is
presented in Table 1.

\begin{table}[h!]
	\small
	\centering
	\begin{tabular}{|ll|ll|ll|}
		\hline
		\multicolumn{2}{|l|}{\textbf{Variable}}                     & \multicolumn{2}{l|}{\textbf{Novices}} & \multicolumn{2}{l|}{\textbf{Experts}} \\ \hline \hline
		\multicolumn{6}{|c|}{\textbf{General Information}}                                                                                                            \\ \hline \hline
		\textbf{Number}        &                                    & 205                    &                       & 40                    &                        \\ \cline{1-6}
		\textbf{Age}           & \textbf{(SD)}                      & 15.20                  & (.95)                 & 27.30                 & (6.33)                 \\ \cline{1-6}
		\textbf{Gender}        & \textbf{(f / m)}                     & 98 / 107                    &                   & 9 / 31                   &                      \\ \hline \hline
		\multicolumn{6}{|c|}{\textbf{Experience in Process Modeling}}                                                                                                 \\ \hline \hline
		\textbf{Analyzed}  & \multicolumn{1}{l|}{\textbf{(SD)}} & \multicolumn{2}{c|}{\multirow{3}{*}{\textbf{\Large NA}}}         & 24.43                 & (21.39)                \\ \cline{1-2} \cline{5-6}
		\textbf{Created}   & \multicolumn{1}{l|}{\textbf{(SD)}} & \multicolumn{2}{l|}{}                          & 19.48                 & (20.33)                \\ \cline{1-2} \cline{5-6}
		\textbf{Activities} & \multicolumn{1}{l|}{\textbf{(SD)}} & \multicolumn{2}{l|}{}                          & 15.70                 & (7.69)                 \\ \hline \hline
		\multicolumn{6}{|c|}{\textbf{Level of Semantics}}                                                                                                              \\ \hline \hline
		\textbf{Concrete}      & \multicolumn{1}{l|}{}              & \multicolumn{2}{l|}{75}                        & \multicolumn{2}{l|}{14}                        \\ \cline{1-6} 
		\textbf{Abstract}      & \multicolumn{1}{l|}{}              & \multicolumn{2}{l|}{58}                        & \multicolumn{2}{l|}{14}                        \\ \cline{1-6} 
		\textbf{Pseudo}        & \multicolumn{1}{l|}{}              & \multicolumn{2}{l|}{72}                        & \multicolumn{2}{l|}{12}                        \\ \cline{1-6} 
	\end{tabular}
\caption{Sample Descriptions for Novices and Experts}
\end{table}

\subsection{Descriptive Statistics}

Table 2 presents the \textit{mean (M)} as well as the \textit{standard deviation
(SD)} for all values obtained by novices (i.e., pupils) and experts. For each
\textit{level of complexity} (i.e., \textit{easy}, \textit{medium}, and
\textit{hard}) and each \textit{level of semantics} (i.e., \textit{concrete},
\textit{abstract}, and \textit{pseudo}), the \textit{achieved score} in
correctly selecting the right statements as well as the \textit{duration needed
(in s)} to solve the task are shown in Table 2. Note that only the
identification of a correct statement result in a point (i.e., two points are
the maximum for each \textit{level of complexity}).

Figs. 5 - 10 depict descriptive \textit{data (means)} of novices (i.e., pupils)
and experts with corresponding \textit{standard
error}\footnote[3]{\scriptsize{}Standard error is used to estimate the standard
deviation of a sampling distribution}. When juxtaposing the results shown in
Figs. 5 and 6, they present the achieved \textit{score} in identifying the
correct statements. Instead of an expected steady decrease of the \textit{score}
with rising \textit{level of complexity}, the \textit{score} for the
\textit{medium} process model is erratic for both samples.

\begin{table}[h!]
	\small
	\centering
	\begin{tabular}{lll|rr|rr|rr|}
		\cline{4-9}
		&                                                       &               & \multicolumn{6}{c|}{\textbf{Version}}                   \\ \cline{2-9} 
		\multicolumn{1}{l|}{}                    & \multicolumn{2}{l|}{\textbf{Indep. \& Dep. Var.}}               & \multicolumn{2}{l|}{\textbf{Concrete M (SD)}}   & \multicolumn{2}{l|}{\textbf{Abstract M (SD)}} & \multicolumn{2}{l|}{\textbf{Pseudo M( SD)}} \\ \hline
		\multicolumn{1}{|l|}{\multirow{6}{*}{\rotatebox{90}{\textbf{Novices}}}} & \multicolumn{1}{l|}{\multirow{2}{*}{\textbf{Easy}}}   & \textbf{Score}  &      1.04 &(.68)             &         1.62& (.64)          & 1.13& (.60)                \\ \cline{3-9} 
		\multicolumn{1}{|l|}{}                   & \multicolumn{1}{l|}{}                                 & \textbf{Duration} &         2282.60 &(821.78)          &   1742.00& (651.41)                &     3474.19 &(1877.23)            \\ \cline{2-9} 
		\multicolumn{1}{|l|}{}                   & \multicolumn{1}{l|}{\multirow{2}{*}{\textbf{Medium}}} & \textbf{Score}  &       .88 &(.66)            &  .53 &(.60)                 &  .63& (.70)               \\ \cline{3-9} 
		\multicolumn{1}{|l|}{}                   & \multicolumn{1}{l|}{}                                 & \textbf{Duration} &          2410.03 &(1019.28)         &   2238.98 &(1308.19)                &    2475 &(1618.57)             \\ \cline{2-9}
		\multicolumn{1}{|l|}{}                   & \multicolumn{1}{l|}{\multirow{2}{*}{\textbf{Hard}}}   & \textbf{Score}  &       1.04& (.73)            &     1.19 &(.58)              &   .82& (.61)              \\ \cline{3-9} 
		\multicolumn{1}{|l|}{}                   & \multicolumn{1}{l|}{}                                 & \textbf{Duration} &          2582.92 & (1729.61)        &    2516.57 & (2123.66)               &  2331.37 & (2154.54)               \\ \hline \hline
		\multicolumn{1}{|l|}{\multirow{6}{*}{\rotatebox{90}{\textbf{Experts}}}} & \multicolumn{1}{l|}{\multirow{2}{*}{\textbf{Easy}}}   & \textbf{Score}  &     1.43 & (.51)            &  1.93 & (.27)     &      1.17 & (.58)     \\ \cline{3-9}  
		\multicolumn{1}{|l|}{}                   & \multicolumn{1}{l|}{}                                 & \textbf{Duration} &     2360.86 & (668.82)      &  2061.57 & (642.45)     &      3656.42 & (1110.17)              \\ \cline{2-9}
		\multicolumn{1}{|l|}{}                   & \multicolumn{1}{l|}{\multirow{2}{*}{\textbf{Medium}}} & \textbf{Score}  &      1.43 & (.65)    &  1.64 & (.63)     &      1.58 &  (.52)              \\ \cline{3-9} 
		\multicolumn{1}{|l|}{}                   & \multicolumn{1}{l|}{}                                 & \textbf{Duration} &    2177.71 & (803.76)  &    2255.14 & (599.44)   &     3379.42 & (954.74)             \\ \cline{2-9}
		\multicolumn{1}{|l|}{}                   & \multicolumn{1}{l|}{\multirow{2}{*}{\textbf{Hard}}}   & \textbf{Score}  &     1.36&  (.48)   &  1.64 & (.48)     &        1.17 & (.72)           \\ \cline{3-9} 
		\multicolumn{1}{|l|}{}                   & \multicolumn{1}{l|}{}                                 & \textbf{Duration} &     4016.57 & (1495.65)    &   4079.07&  (1475.68)    &     6755.25&  (2384.94)                \\ \hline
	\end{tabular}
	\caption{Descriptive Results for Novices and Experts}
\end{table}

\begin{figure}[h!]
	\begin{minipage}{0.45\textwidth} 
		\includegraphics[width=\linewidth]{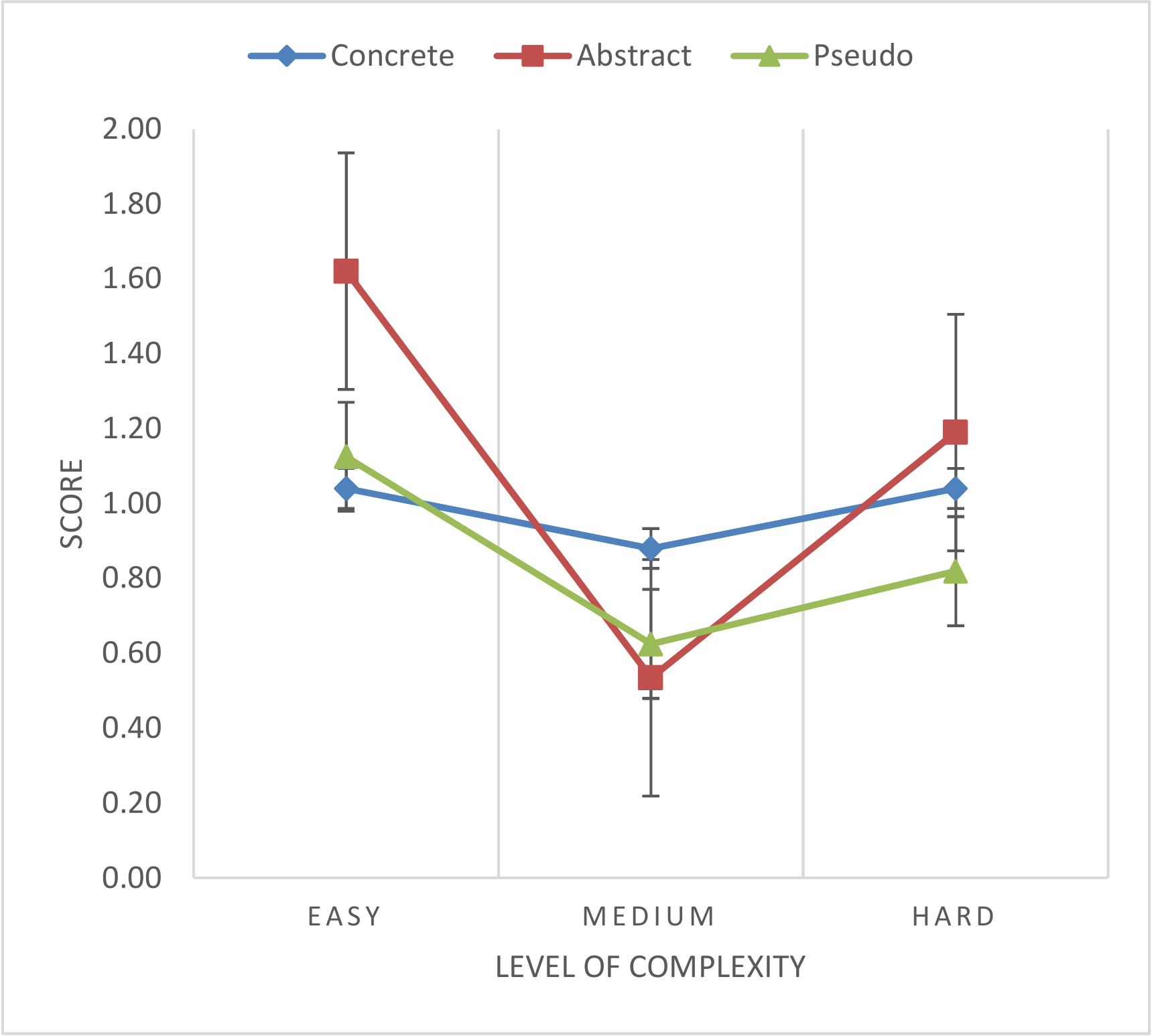}
		\caption{Score Achieved by Novices}
	\end{minipage}
	\hfill
	\begin{minipage}{0.45\textwidth}
		\includegraphics[width=\linewidth]{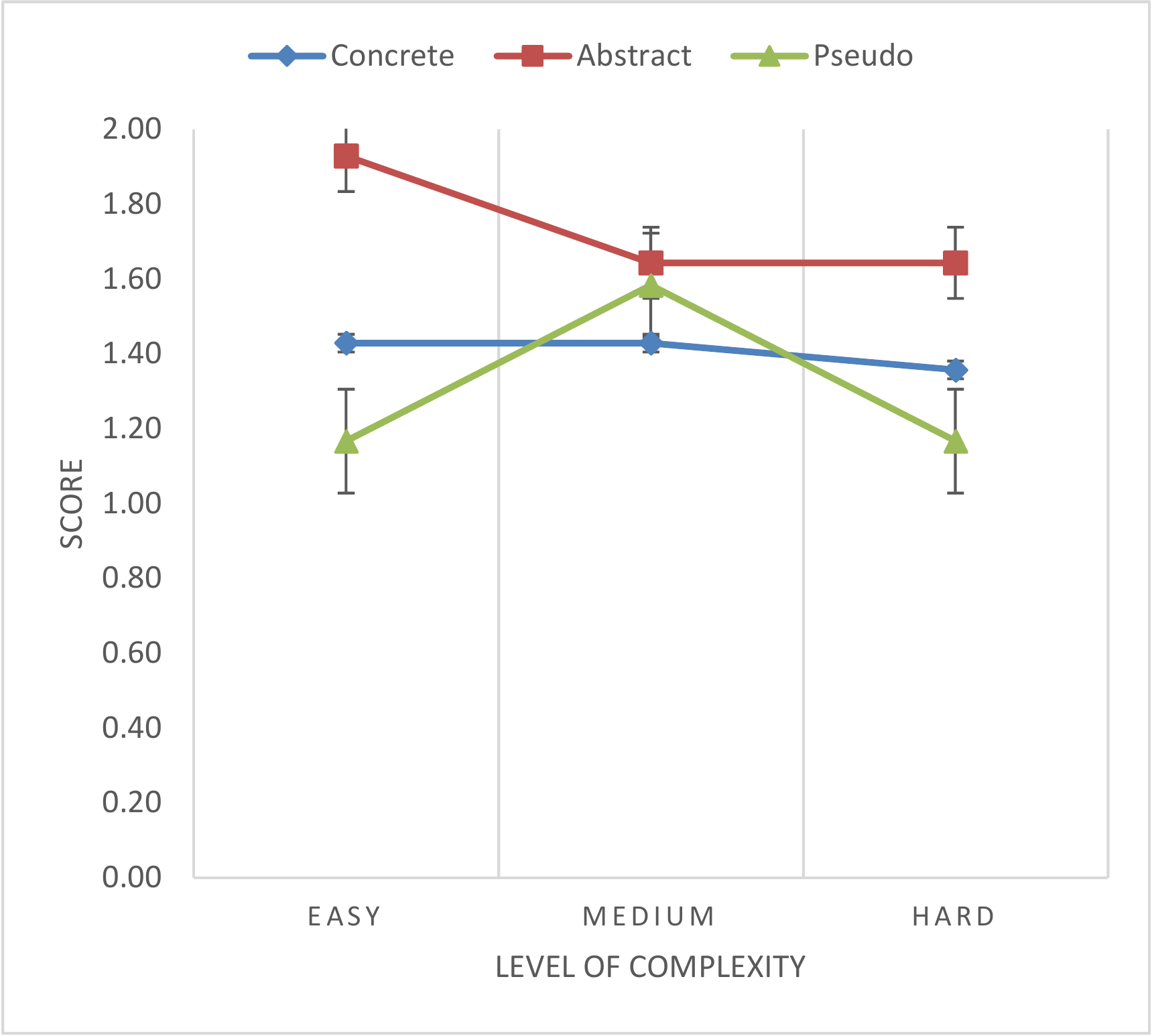}
		\caption{Score Achieved by Experts}
	\end{minipage}
\end{figure}

Regarding the \textit{duration} needed by pupils for corresponding process
models (cf. Fig. 7), a clear difference between the different \textit{levels of
semantics} can be observed in the easy process model. However, the
\textit{duration} to solve a task reaches the same value with increasing
\textit{level of complexity}. For experts (cf. Fig. 8), tasks with process
models reflecting an \textit{abstract} and \textit{concrete level of semantics}
are solved faster than the \textit{pseudo} ones.\\
As depicted in Fig. 9, experts
show a better performance in comprehending process models compared to the
novices (i.e., pupils). According to Fig. 10, interestingly process models
showing an \textit{abstract level of semantics} are understood easier than
models with a \textit{concrete} or \textit{pseudo level of semantics}, thus
\textit{abstract} labels having a positive effect on process model
comprehension. However, these observations are merely based on descriptive
statistics. For a more rigid investigation, dependent variables are tested for
statistical significance.

\begin{figure}[h!]
	\centering
	\begin{minipage}{0.45\textwidth}
		\includegraphics[width=\linewidth]{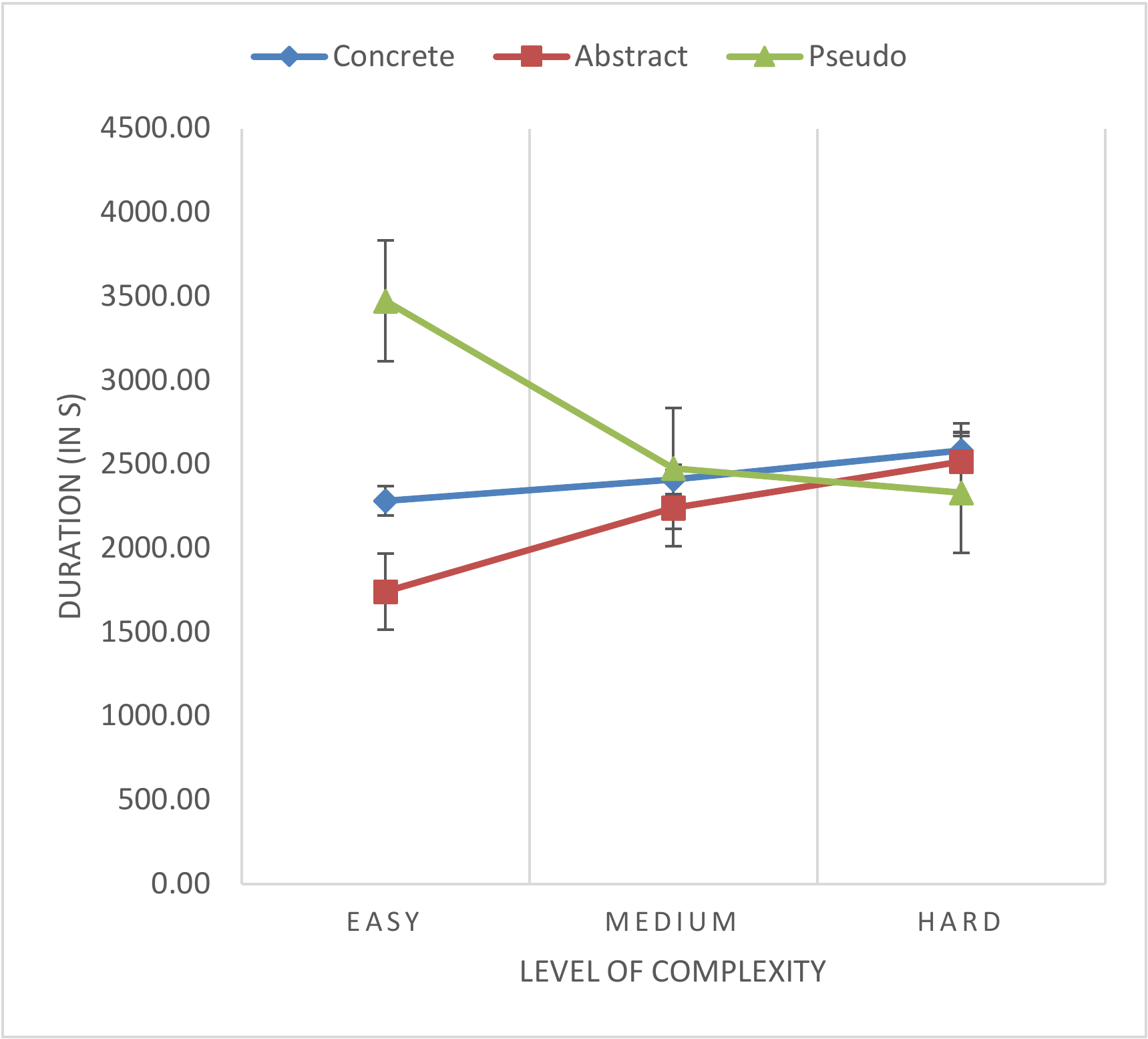}
		\caption{Duration Needed by Novices}
	\end{minipage}
\hfill
	\begin{minipage}{0.45\textwidth}
		\includegraphics[width=\linewidth]{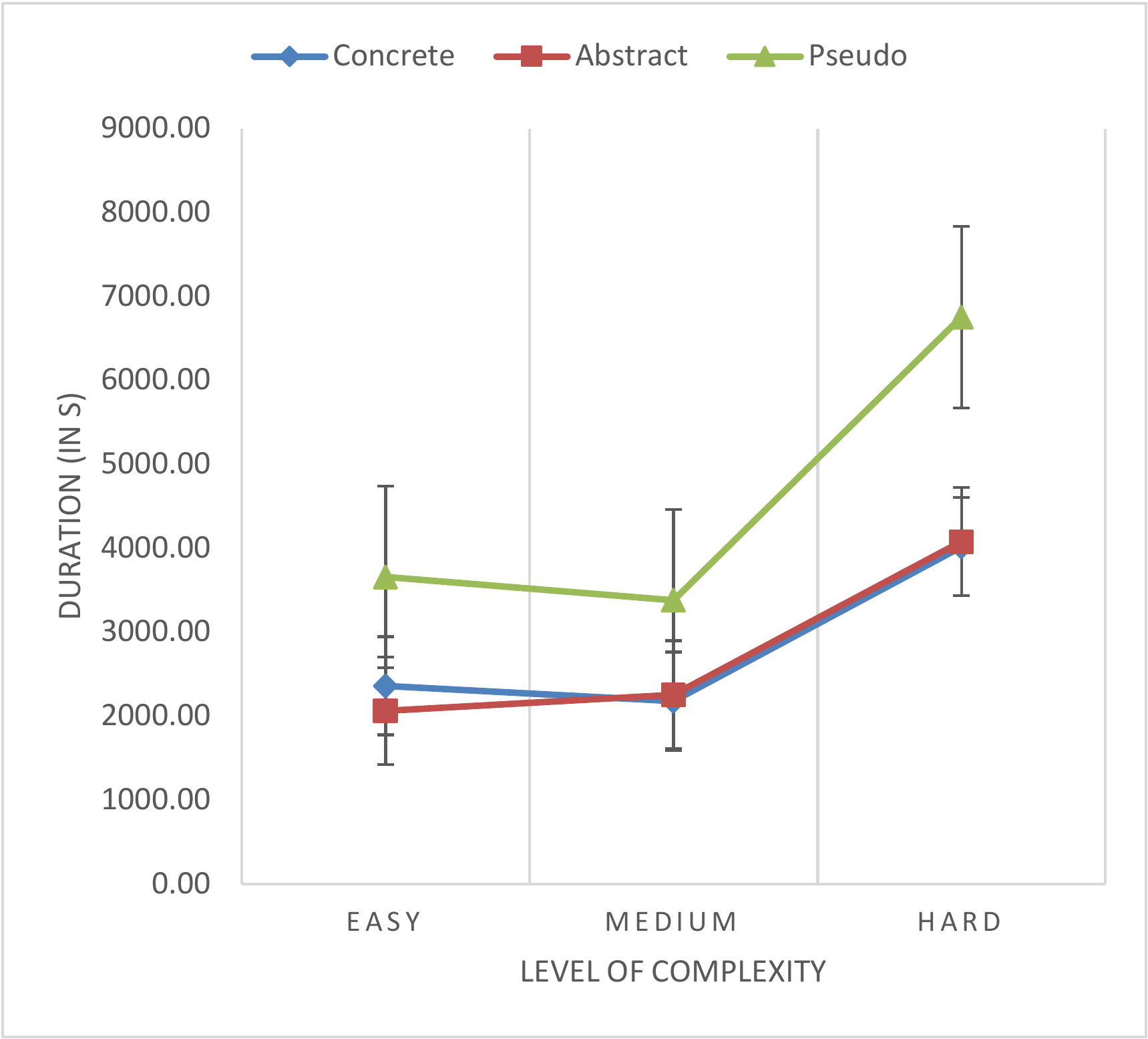}
		\caption{Duration Needed by Experts}
	\end{minipage}
\end{figure}

\begin{figure}[h!]
	\centering
	\begin{minipage}{0.45\textwidth}
	\includegraphics[width=\linewidth]{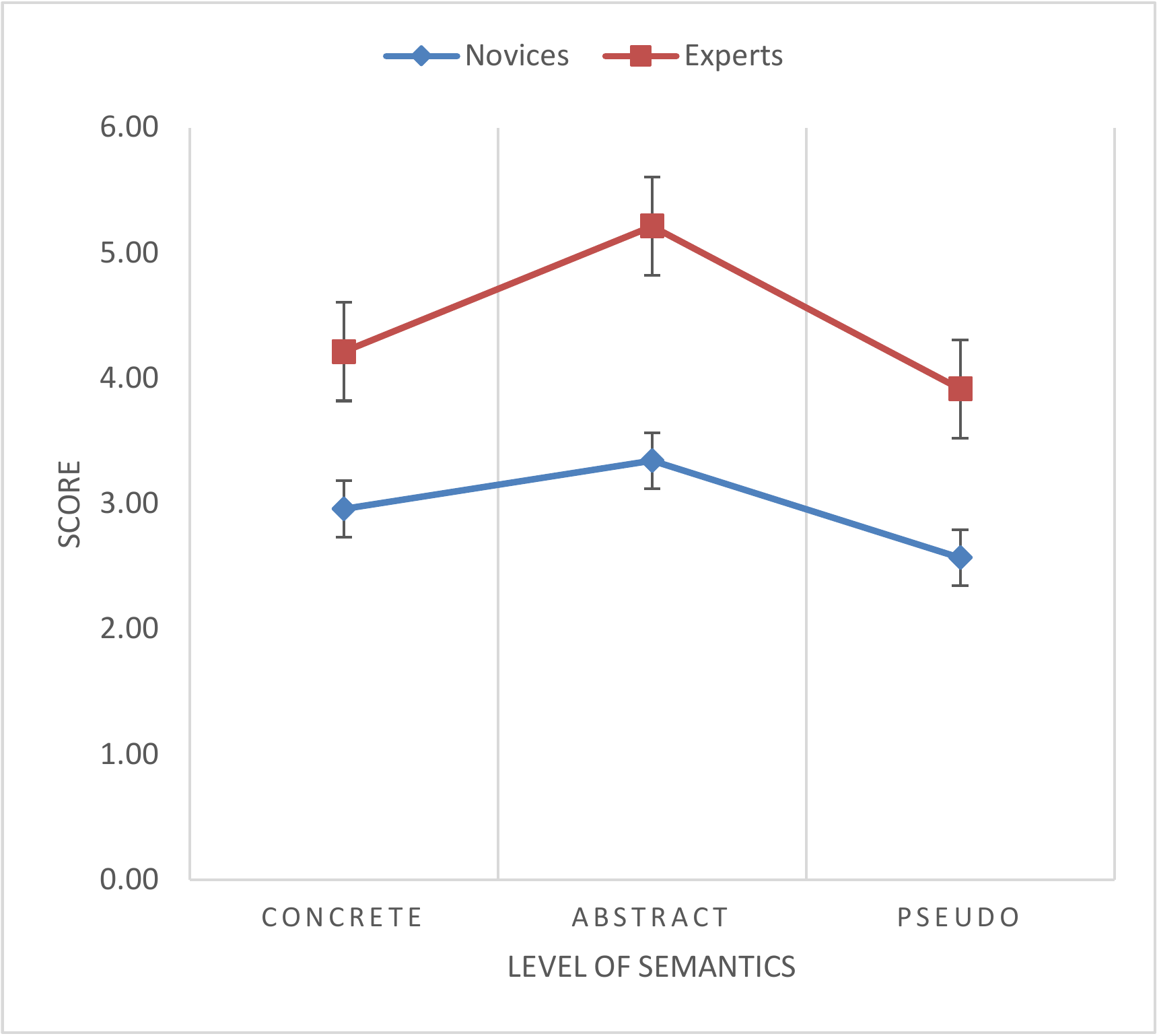}
	\caption{Score for both Samples}
\end{minipage}
\hfill
\begin{minipage}{0.45\textwidth}
	\includegraphics[width=\linewidth]{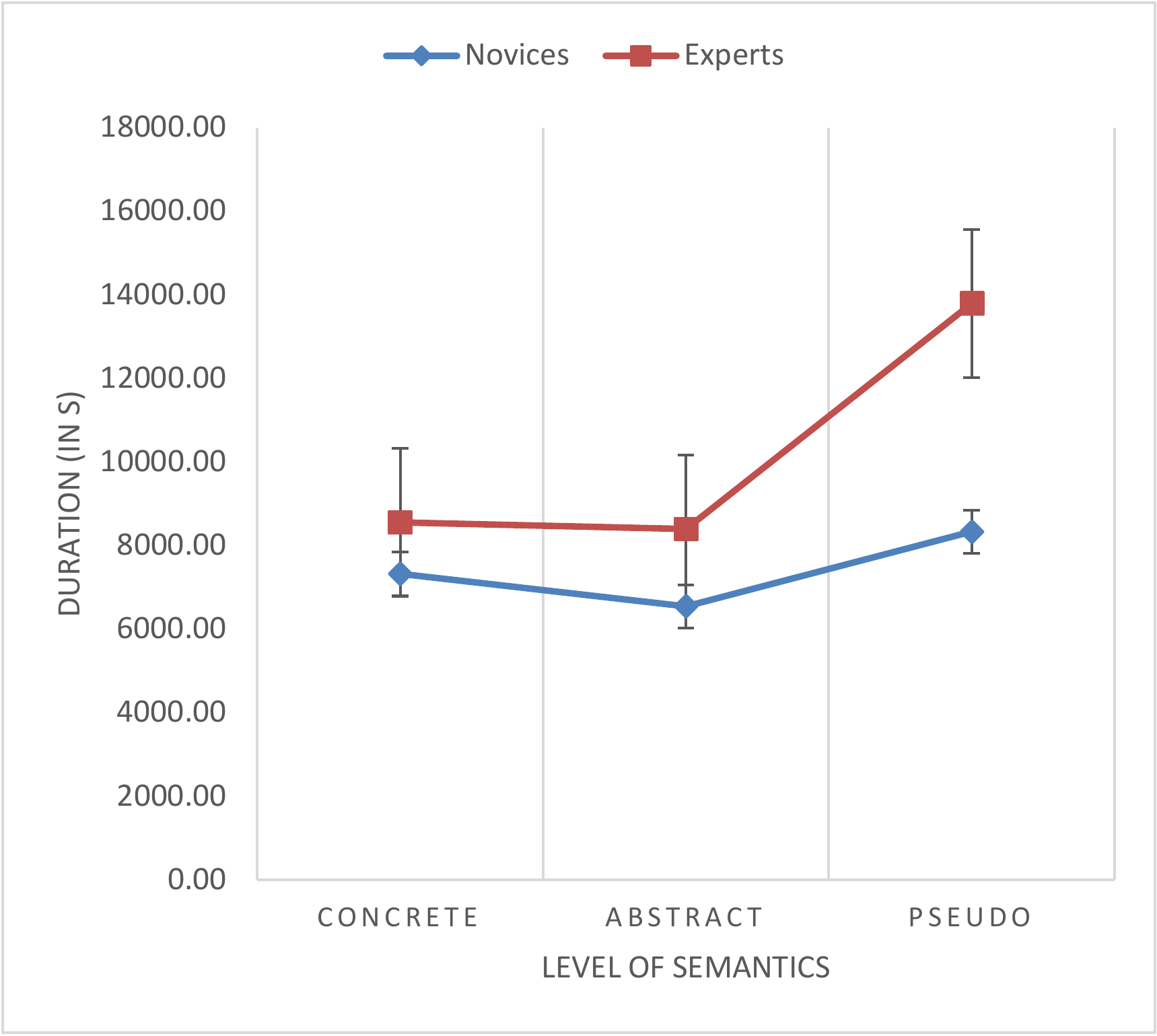}
	\caption{Duration for both Samples}
\end{minipage}
\end{figure}

\subsection{Inferential Statistics}

The \textit{analysis of variance (ANOVA)} for repeated measurements is performed
for each dependent variable (i.e., \textit{score} and \textit{duration}).
Thereby, \textit{main effects}\footnote[4]{\scriptsize{}ME is the effect of the
variable averaging over all levels of the same variable.} \textit{(ME)} and
\textit{interaction effects}\footnote[5]{\scriptsize{}IE measures the
interaction of a variable with another variable(s).} \textit{(IE)} are considered. The main
effects for \textit{level of complexity (ME 1)}, \textit{level of semantics (ME
2)}, and \textit{level of expertise (ME 3)} are investigated (cf. Sect 3.2).
Furthermore, the following interaction effects are analyzed:
\textit{complexity*semantics (IE 1)}, \textit{complexity*expertise (IE 2)},
\textit{semantics*expertise (IE 3)}, and \textit{complexity*semantics*expertise
(IE 4)}. All statistical tests are performed two-tailed with the significance
value being set to $p < .05$. Table 3 presents the results calculated for
\textit{both samples} as well as for \textit{novices (n))} (i.e.,
\textit{pupils}) and \textit{experts (e)} being considered separately.

\begin{table}[h]
	\small
	\centering
	\begin{tabular}{ll|ll|ll|}
\cline{3-6}
&        & \multicolumn{4}{c|}{\textbf{Dependent Variables}}                                       \\ \cline{2-6} 
\multicolumn{1}{l|}{} 		& \textbf{Effect} & \multicolumn{2}{c|}{\textbf{Score}}            & \multicolumn{2}{c|}{\textbf{Duration}}         \\ \hline
		\multicolumn{1}{|l|}{\multirow{5}{*}{\rotatebox[]{90}{\textbf{Both}}}} & \textbf{ME 1}   & $F(1.90;475.32)=7.32;$  & $p<.01$ & $F(1.62;386.44)=49.81;$ & $p<.01$ \\ \cline{2-6}
		\multicolumn{1}{|l|}{}                  & \textbf{ME 2}  & $F(2; 239)=7.83;$       & $p<.01$ & $F(2;239)=12.04;$       & $p<.01$\\ \cline{2-6}
		\multicolumn{1}{|l|}{}                  & \textbf{ME 3}   & $F(1;239)=47.54;$       & $p<.01$ & $F(1;239)=20.71;$       & $p<.01$ \\ \cline{2-6}
		\multicolumn{1}{|l|}{}                  & \textbf{IE 1}   & $F(3.98;475.32)=4.47;$  & $p<.01$ & $F(3.23;386.44)=2.46;$  & $p=.06$         \\ \cline{2-6}
		\multicolumn{1}{|l|}{}                  & \textbf{IE 2}  & $F(1.90;475.32)=10.61;$ & $p<.01$ & $F(1.62;386.44)=46.20;$ & $p<.01$ \\ \cline{2-6}
		\multicolumn{1}{|l|}{}                  & \textbf{IE 3}  & $F(1.90;475.32)=10.61;$ & $p<.01$ & $F(1.66;386.44)=2.46;$ & $p<.01$ \\ \cline{2-6}
		\multicolumn{1}{|l|}{}                  & \textbf{IE 4}  & $F(3.98;475.32)=1.41;$ & $p=.23$ & $F(3.23;386.44)=6.47;$ & $p<.01$ \\ \hline \hline
		\multicolumn{1}{|l|}{\multirow{3}{*}{\rotatebox[]{90}{\textbf{Novice}}}} & \textbf{ME\textsubscript{n} 1}   & $F(1.98;399.79)=47.98;$ & $p<.01$ & $F(1.55;57.46)=79.97;$ & $p<.01$ \\ \cline{2-6}
		\multicolumn{1}{|l|}{}                  & \textbf{ME\textsubscript{n} 2}  & $F(2;202)=6.02;$ & $p<.01$     & $F(2;37)=14.36;$ & $p<.01$ \\ \cline{2-6}
		\multicolumn{1}{|l|}{}                  & \textbf{IE\textsubscript{n} 1}  & $F(3.96;399.79)=10.30;$ & $p<.01$ & $F(3.11;57.46)=3.41;$ & $p=.02$ \\ \hline \hline
		\multicolumn{1}{|l|}{\multirow{3}{*}{\rotatebox[]{90}{\textbf{Expert}}}} & \textbf{ME\textsubscript{e} 1}   & $F(1.96;72.54)=1.13;$ & $p=.33$  & F(1.55;57.46)=79.97; & $p<.01$ \\ \cline{2-6}
		\multicolumn{1}{|l|}{}                  & \textbf{ME\textsubscript{e} 2}  & $F(2;37)=5.01;$ & $p=.01$     & $F(2;37)=14.36;$ & $p<.01$\\ \cline{2-6}
		\multicolumn{1}{|l|}{}                  & \textbf{IE\textsubscript{e} 1}  & $F(3.92;72.54)=1.69;$ & $p=.16$ & $F(3.11;57.46)=3.41;$ & $p=.02$ \\ \hline
	\end{tabular}
	\caption{Results from Analysis of Variance (ANOVA)}
\end{table}

In summary, for both samples, the statistical analyses show a high significance
for all variables, except the value for \textit{IE 4} regarding the
\textit{score} and \textit{IE 1} regarding the \textit{duration}. Thereby, the
latter almost reaches statistical significance. Concerning the \textit{score}
achieved by novices, there are significant differences between the \textit{level
of complexity (ME\textsubscript{n} 1)} and the \textit{level of semantics
(ME\textsubscript{n} 2)}. Consequently, the interaction effect between these two
variables reaches statistical significance \textit{(IE\textsubscript{n} 1)}.
Concerning the \textit{duration}, there is no significant difference regarding
the \textit{level of complexity (ME\textsubscript{n} 1)}. By contrast, a
statistical significance related to the \textit{level of semantics
(ME\textsubscript{n} 2)} is measurable. Consequently, the interaction effect
shows a statistically significant result \textit{(IE\textsubscript{n} 1)}.
Regarding the experts' \textit{score}, there is no significant difference
regarding the \textit{\textit{level of complexity} (ME\textsubscript{e} 1)}, but
statistical significance is observable between the \textit{level of semantics
(ME\textsubscript{e} 2)}. Furthermore, the interaction effect shows no
significant difference \textit{(IE\textsubscript{e} 1)}. Considering the
\textit{duration}, there are significant differences between the
\textit{\textit{level of complexity} (ME\textsubscript{e} 1)} as well as the
\textit{level of semantics (ME\textsubscript{e} 2)} and, hence, likewise the
interaction effect reflects a significant difference
\textit{(IE\textsubscript{e} 1)}.

\subsection{Discussion}
According to the study results, pupils do not outperform experts regarding
process model comprehension \textit{(RQ 3)}. To be more precise, the modeling
experts achieved better results than the pupils (i.e., novices). However, as
indicated by descriptive statistics (cf. Sec. 4.2), pupils are able to
comprehend process models correctly \textit{(RQ 2 + RQ 3)}. Concerning the
\textit{two-out-of-four} combination, pupils are able to identify at least one
correct statement per average \textit{(RQ 2)}.\\
Regarding the \textit{levels of semantics}, interestingly process models with an
\textit{abstract labeling} are comprehended better compared to models with
\textit{concrete} or \textit{pseudo labeling} \textit{(RQ 3)}. This can be
explained by the fact that pupils do not need to cope with parsing the relevant
\textit{semantic} and \textit{pragmatic information} in the respective process
models. Due to the \textit{abstract labeling}, the cognitive load caused by the
\textit{semantics} as well as \textit{pragmatics} can be neglected. This leads
to more increased capacity in the working memory and enables us to solely focus
on process model \textit{syntactics} and behavior. Consequently, the probability
for errors is inferior \cite{vanderfeesten2007quality}. \\
Concerning the \textit{concrete level of semantics}, during process model
comprehension, pupils seem to consider their own experiences with the respective
process scenario. In particular, they try to match the given statements with
their own process reflection of the scenario in their minds and, therefore,
answer related statements based on syllogisms, though the answers might be
false. As \textit{pseudowords} reflect no lexical semantics, they appear to be
an additional challenge for pupils. A clear difference with respect to the
\textit{duration} needed to comprehend a process model can be observed for the
\textit{easy} process models, the \textit{comprehension duration} reaches the
same value for all \textit{levels of semantics} with increasing \textit{level of
complexity} \textit{(RQ 2 + RQ 3)}. Note that with a \textit{pseudo} labeling, a
decrease of the \textit{duration} needed can be observed. However, the same
effect cannot be observed for the \textit{score} achieved. This can be explained
by the fact that a learning effect occurs between the \textit{levels of
complexity}, leading to a faster comprehension of respective process models
\cite{zimoch2017using}.  \\
Concerning the \textit{medium} process model, a particular phenomenon is
discernible (cf. Figs. 5 and 6). The pupils showed a significant decrease in the
\textit{score}, while the \textit{score} increases for the \textit{hard} process
model. In turn, experts achieved a slightly better result regarding the
\textit{score} for the \textit{medium} process model, but showed a decrease for
the \textit{hard} model. One might anticipate that the \textit{scores} will
decrease with rising \textit{level of complexity}. \\
Having a closer look at the process model with a \textit{medium level of
complexity}, one can see that the model comprises a parallel path (i.e., AND
gateway) as well as a loop (i.e., XOR gateway). Moreover, the two correct
statements associated with this model refer to the interpretation of the
parallel path and the loop. According to the results, pupils seem to experience
difficulties in the correct interpretation of the parallel path. The loop, in
turn, is comprehended correctly by them. The same effect can be observed in the
subsequent task related to the comprehension of the \textit{hard} process model.
The \textit{hard} process model comprises a parallel path, a loop, and two
decision points (i.e., XOR gateway). The statements in this model refer on the
interpretation of the XOR gateways (i.e., loop and decision points). Pupils are
able to interpret and comprehend the loop as well as the decision points
correctly, which does not fully apply to the parallel path. This reconfirms
observations we made in a prior study, which revealed that the first gateway
appearing along the reading direction seems to be more challenging to comprehend
compared to the subsequent ones \cite{zimoch2}. A common approach for
interpreting a gateway is to consider the process scenario in more detail.
This way, the behavior of the remaining gateways can be derived. \\
Regarding the \textit{scores} achieved by the experts and they \textit{duration}
needed, it has become evident that process models with an \textit{abstract level
of semantics} are comprehended easier compared to models with \textit{concrete}
and \textit{pseudo} labels. Moreover, process models with an \textit{abstract
level of semantics} are comprehended fastest, whereas for the other two
\textit{levels of semantics}, the \textit{duration} needed to comprehend
respective process models is approximately the same. Experts achieved a
considerably better \textit{score} in identifying the correct statements
compared to pupils. However, experts need more \textit{time} for comprehending a
process model. As a reason for this phenomenon, experts have spent more time in
parsing information on the syntactics and semantics of a process model. \\
Our findings might have several implications. The conducted study provides
information on how novices comprehend process models compared to experts.
Interestingly, the use of an \textit{abstract level of semantics} fosters
learning of process modeling notations, e.g., reducing the complexity for
parsing the semantic information of a process model and, hence, the focus can be
set entirely on the syntactical interpretation of the modeling language. This
insight might be useful for the analysis of process models which are
syntactically not sound. In general, process models can be read and comprehended
intuitively. On the other hand, particular modeling constructs (e.g., AND
gateway) seem to be more challenging to comprehend compared to others (e.g., XOR
gateway). This indicates to focus on modeling constructs and their respective
behavior, which are likely to be more difficult to comprehend. Moreover, based
on our findings, modeling guidelines can be derived towards creating better
comprehensible process models.


\subsection{Threats to Validity}
Although validity factors are carefully considered, there are threats to
validity that need to be discussed. \textit{First}, the use of two different
procedures underlying the study for pupils and experts limit its validity. The
procedure for pupils takes significantly more completion time ($\sim$ 40
minutes) compared to the one of experts ($\sim$ 10 minutes). Hence, this might
have a negative impact on the outcome, due to particular state of minds (e.g.,
tiredness, boredom). \textit{Second}, participants could discovered that two
statements are always true. As a consequence, they know that always two
statements need to be selected.
\textit{Third}, the respective \textit{level of complexity} reflected by the
process models constitutes another threat to validity. The process models might
be considerably unbalanced between the \textit{levels of complexity}. In detail,
working memory capacity of participants may be exceeded. The same might be
applied to the different \textit{levels of semantics} as well as the single
statements related to the process models. \textit{Fourth}, as another risk, no
professionals from industry are involved, but prospective ones (i.e., students).
Although various investigations have shown that students are proper substitutes
for professionals in empirical studies \cite{Host2000}, results for
professionals might differ. \textit{Fifth}, the representativeness of the
results is limited due to the relatively small sample of experts ($n=40$),
although this number is rather higher compared to similar studies. Accordingly,
the different sample sizes between pupils ($n=205$) and experts ($n=40$) make it
more probable to detect significant results for pupils than for experts. Note
that we currently address these limitations in other studies to obtain more
accurate results allowing for a further generalization.

\section{Related Work}

Research on process model comprehension can be classified into
\textit{subjective} and \textit{objective comprehensibility}. \\
Regarding \textit{subjective comprehensibility}, \cite{Mendling2018} gives
insights into various characteristics (e.g., theoretical knowledge) of an
individual that influence process model comprehension. In turn,
\cite{khatri2006understanding} confirms that process scenarios from a familiar
application domain represents a key factor for understanding conceptual models.
Finally, \cite{burattin2017detection} presents a study focusing on visual
features of process models (i.e., flow consistency) and their impact on human
perception of process models. \\
Regarding \textit{objective comprehensibility}, \cite{CORRADINI2018129} provides
fifty guidelines for improving BPMN 2.0 process models with respect to their
comprehensibility. A systematic literature review of the factors influencing the
comprehension of process models is presented in \cite{DIKICI2018112}. Moreover,
\cite{wang2017effect} presents an experiment investigates the effects of
integrating business rules into a process model. \\
The empirical study in \cite{Figl2011} shows that subjects who are confronted
with complex process models quickly encounter cognitive limitations, which
impairs process model comprehension. Finally, \cite{HouyFL14} demonstrated that
confronting individuals with a cognitive overload will have an adverse effect on
model comprehension. \\
Various works in literature exist investigating how different types of labels
(e.g., \textit{concrete} or \textit{abstract}) are interpreted, understood, and
processed by individuals. For example, \cite{price1996demonstrating}
demonstrates that reading pseudowords results in a higher cognitive load of the
working memory, having a negative impact on the performance in processing
respective tasks. In turn, \cite{word} showed that letters, words, and simple
text, with or without context, are read at different speeds based on a set of
individual differences (e.g., intelligence). \\
Regarding the labeling of activities in process modeling,
\cite{mendling2010activity} presents different practices for labeling activities
and examines their usability. Based on the insights obtained from this study,
specific labeling styles are recommended for process modeling. In addition, the
quality of activity labels is addressed in \cite{leopold2012refactoring} and a
technique for refactoring activity labels is presented. Finally, the visual
design of element labels in a process model is addressed in
\cite{koschmider2015comprehensive}, providing recommendations for the design of
element labels in process models. \\
It is known that children as well as adolescents perceive and interpret their
surroundings differently compared to grown-ups, e.g., \cite{punch2002research}
discusses issues when comparing children with grown-ups and, hence, why they
have to be considered differently in research. Finally, \cite{rudduck2000pupil}
suggests to tackle research from a pupil's perspective in order to unravel new
learning strategies. However, to the best of our knowledge, so far no approach
has investigated the influence of a leaner's visual literacy in cultural
education on the comprehension of process models

\section{Summary and Outlook}

This paper investigated whether novices (i.e., pupils) are able to comprehend
business process models, although they have no previous knowledge on process
model notations. In total, $n=205$ pupils had to solve a visual
task related to process model comprehension. As a benchmark, we performed a
similar study with $n=40$ process modeling experts. Although experts
outperformed pupils in the respective comprehension tasks, the results indicate
that process models can be properly understood by pupils as well. Thereby,
process models with an \textit{abstract level of semantics} are easier to
comprehend compared to process models with a \textit{concrete} or \textit{pseudo
level of semantics}.
Based on these findings, one may conclude that using an \textit{abstract}
labeling of elements fosters the learning of process modeling notations.
Moreover, \textit{abstract} labels assist in the analysis of syntactically
unsound process models. Our study provides empirical evidence to focus on
process modeling constructs that are likely to be difficult to comprehend (e.g.,
AND gateway). Our insights suggest that it might be beneficial to provide
additional guidance in reading such constructs. Furthermore, modeling guidelines
(e.g., 7PMG \cite{Mendling2010}) for the creation of better comprehensible
process models can be provided based on the findings derived from the conducted
study as well as existing modeling tools can be enhanced with supplementary
features. \\
Although our results reveal interesting insights, further research is required.
Therefore, we will conduct more process model comprehension studies with pupils
from higher classes as well as teachers, students, and graduates (i.e.,
learners) to investigate the effects of cultural education (e.g., educational
years) on process model comprehension. Furthermore, we will analyze data
obtained from eye tracking and compare the applied strategies of learners to
read and comprehend a process model with the one of experts.
Finally, we will consider the influence of visual literacy and cultural
education of learners with respect to business process models
in more detail. This might reveal insights on how to foster the comprehension of
process models.

\bibliographystyle{splncs} 
\bibliography{BPM_2018}

\end{document}